\documentclass[twocolumn,showpacs,aps,prd,superscriptaddress]{revtex4}
\usepackage{graphicx}
\usepackage{dcolumn}
\usepackage{amsmath}
\usepackage{epsfig}
\usepackage{relsize}
\usepackage{sidecap}
\usepackage{graphicx}

\RequirePackage{xspace}

\def\pep2{PEP-II}

\newcommand{\mev}{MeV\xspace}
\newcommand{\gevc}{GeV\!/\ensuremath{\ensuremath{c}}\xspace}

\newcommand{\gevcc}{GeV\!/\ensuremath{\ensuremath{c^2}}\xspace}
\newcommand{\mevcc}{MeV\!/\ensuremath{\ensuremath{c^2}}\xspace}

\newcommand{\MeV}{\mev }

\newcommand{\MeVcc}{\mevcc}

\newcommand{\GeVcc}{\gevcc }

\def\etal     {{\it et\,al.}}

\def\babar{\mbox{\slshape B\kern-0.1em{\smaller A}\kern-0.1em B\kern-0.1em{\smaller A\kern-0.2em R}}}

\def\invfb    {fb\ensuremath{^{-1}}\xspace}

\def\Dz       {\ensuremath{D^0}\xspace}
\def\Dbar     {\kern 0.2em\overline{\kern -0.2em D}{}\xspace}

\def\Dp       {\ensuremath{D^+}\xspace}

\def\epem     {\ensuremath{e^+e^-}\xspace}

\def\piz      {\ensuremath{\pi^0}\xspace}
\def\pip      {\ensuremath{\pi^+}\xspace}
\def\pim      {\ensuremath{\pi^-}\xspace}

\def\Km       {\ensuremath{K^-}\xspace}

\def\DToKPi  {\ensuremath{D^0 \rightarrow K^-\pi^+}\xspace }
\def\KPi  {\ensuremath{K^-\pi^+}\xspace }
\def\DToKPiPiPi  {\ensuremath{D^0 \rightarrow K^- \pi^+ \pi^- \pi^+}\xspace}

\def\DCToKPiPi  {\ensuremath{D^+ \rightarrow K^- \pi^+ \pi^+}\xspace}
\def\KPiPi  {\ensuremath{ K^- \pi^+ \pi^+}\xspace}

\def\Dstar {\ensuremath{D^{*+}}\xspace}
\def\Dstarz {\ensuremath{D^{*0}}\xspace}
\def\DstarPi {\ensuremath{D^{*+}\pi^{-}}\xspace}
\def\DstarPiWS {\ensuremath{D^{*+}\pi^{+}}\xspace}
\def\DPi {\ensuremath{D^{+}\pi^{-}}\xspace}

\def\DzPi {\ensuremath{D^{0}\pi^{+}}\xspace}

\def\DstarPlusPi0 {\ensuremath{ {D}^{*+}\pi^0}\xspace}
\def\Dst {\ensuremath{D^{*+}}\xspace}
\def\Dstst {\ensuremath{D^{**}}\xspace}
\def\DTwentyFourThirty {\ensuremath{{D}^\prime_1(2430)}\xspace}
\def\DTwentyFourThirtyNeutral {\ensuremath{{D}^\prime_1(2430)^0}\xspace}
\def\DTwentyFourTwenty {\ensuremath{D_1(2420)}\xspace}
\def\DTwentyFourTwentyNeutral {\ensuremath{ D_1(2420)^0}\xspace}
\def\DTwentyFourTwentyCharged {\ensuremath{ D_1(2420)^+}\xspace}
\def\DTwentyFourSixty {\ensuremath{{D}^*_2(2460)}\xspace}
\def\DTwentyFourSixtyNeutral {\ensuremath{ {D}^*_2(2460)^0}\xspace}
\def\DTwentyFourSixtyCharged {\ensuremath{ {D}^*_2(2460)^+}\xspace}
\def\DTwentyFourHundred {\ensuremath{{D}^*_0(2400)}\xspace}

\def\DTwentyFiveFiftyNeutral {\ensuremath{{D}(2550)^0}\xspace}

\def\DTwentySixHundred {\ensuremath{{D^*}(2600)}\xspace}
\def\DTwentySixHundredNeutral {\ensuremath{{D^*}(2600)^0}\xspace}
\def\DTwentySixHundredCharged {\ensuremath{{D^*}(2600)^+}\xspace}

\def\DTwentySevenFiftyNeutral {\ensuremath{{D}(2750)^0}\xspace}

\def\DTwentySevenSixty {\ensuremath{{D^*}(2760)}\xspace}
\def\DTwentySevenSixtyNeutral {\ensuremath{{D^*}(2760)^0}\xspace}
\def\DTwentySevenSixtyCharged {\ensuremath{{D^*}(2760)^+}\xspace}

\newcommand{\cosHelicity}{\ensuremath{ \cos\theta_{H}}\xspace }

\newcommand{\cc}{\ensuremath{c\bar{c}}\xspace}

\def\MDPi {\ensuremath{M(D^+\pi^-)}\xspace}
\def\MDzPi {\ensuremath{M(D^0\pi^+)}\xspace}
\def\MDstarPi {\ensuremath{M(D^{*+}\pi^-)}\xspace}

\newcommand{\DTwentyFourSixtyNYield}{242.8}
\newcommand{\DTwentyFourSixtyNYieldStat}{1.8}
\newcommand{\DTwentyFourSixtyNYieldSyst}{3.4}
\newcommand{\DTwentyFourSixtyNMass}{2462.2}
\newcommand{\DTwentyFourSixtyNMassStat}{0.1}
\newcommand{\DTwentyFourSixtyNMassSyst}{0.8}
\newcommand{\DTwentyFourSixtyNWidth}{50.5}
\newcommand{\DTwentyFourSixtyNWidthStat}{0.6}
\newcommand{\DTwentyFourSixtyNWidthSyst}{0.7}

\newcommand{\DTwentySixHundredNYield}{26.0}
\newcommand{\DTwentySixHundredNYieldStat}{1.4}
\newcommand{\DTwentySixHundredNYieldSyst}{ 6.6}
\newcommand{\DTwentySixHundredNMass}{2608.7}
\newcommand{\DTwentySixHundredNMassStat}{2.4}
\newcommand{\DTwentySixHundredNMassSyst}{2.5}
\newcommand{\DTwentySixHundredNWidth}{93}
\newcommand{\DTwentySixHundredNWidthStat}{6}
\newcommand{\DTwentySixHundredNWidthSyst}{13}
\newcommand{\DTwentySixHundredNSig}{3.9}

\newcommand{\DTwentySevenSixtyNYield}{11.3}
\newcommand{\DTwentySevenSixtyNYieldStat}{0.8}
\newcommand{\DTwentySevenSixtyNYieldSyst}{1.0}
\newcommand{\DTwentySevenSixtyNMass}{2763.3}
\newcommand{\DTwentySevenSixtyNMassStat}{2.3}

\newcommand{\DTwentySevenSixtyNWidth}{60.9}
\newcommand{\DTwentySevenSixtyNWidthStat}{5.1}
\newcommand{\DTwentySevenSixtyNWidthSyst}{3.6}
\newcommand{\DTwentySevenSixtyNSig}{8.9}

\newcommand{\DTwentyFourTwentyNYield}{102.8}
\newcommand{\DTwentyFourTwentyNYieldStat}{1.3}
\newcommand{\DTwentyFourTwentyNYieldSyst}{2.3}
\newcommand{\DTwentyFourTwentyNMass}{2420.1}
\newcommand{\DTwentyFourTwentyNMassStat}{0.1}
\newcommand{\DTwentyFourTwentyNMassSyst}{0.8}
\newcommand{\DTwentyFourTwentyNWidth}{31.4}
\newcommand{\DTwentyFourTwentyNWidthStat}{0.5}
\newcommand{\DTwentyFourTwentyNWidthSyst}{1.3}
\newcommand{\DTwentyFiveFiftyNYield}{34.3}
\newcommand{\DTwentyFiveFiftyNYieldStat}{6.7}
\newcommand{\DTwentyFiveFiftyNYieldSyst}{9.2}
\newcommand{\DTwentyFiveFiftyNMass}{2539.4}
\newcommand{\DTwentyFiveFiftyNMassStat}{4.5}
\newcommand{\DTwentyFiveFiftyNMassSyst}{6.8}
\newcommand{\DTwentyFiveFiftyNWidth}{130}
\newcommand{\DTwentyFiveFiftyNWidthStat}{12}
\newcommand{\DTwentyFiveFiftyNWidthSyst}{13}
\newcommand{\DTwentyFiveFiftyNSig}{3.0}

\newcommand{\DTwentySixHundredNYieldDstPi}{50.2}
\newcommand{\DTwentySixHundredNYieldDstPiStat}{3.0}
\newcommand{\DTwentySixHundredNYieldDstPiSyst}{6.7}
\newcommand{\DTwentySixHundredNDstPiSig}{7.3}

\newcommand{\DTwentyFourTwentyNYieldF}{214.6}
\newcommand{\DTwentyFourTwentyNYieldStatF}{1.2}
\newcommand{\DTwentyFourTwentyNYieldSystF}{6.4}
\newcommand{\DTwentyFourSixtyNYieldDstPi}{136}
\newcommand{\DTwentyFourSixtyNYieldDstPiStat}{2}
\newcommand{\DTwentyFourSixtyNYieldDstPiSyst}{13}
\newcommand{\DTwentyFiveFiftyNYieldF}{98.4}
\newcommand{\DTwentyFiveFiftyNYieldStatF}{8.2}
\newcommand{\DTwentyFiveFiftyNYieldSystF}{38}
\newcommand{\DTwentySixHundredNYieldDstPiF}{71.4}
\newcommand{\DTwentySixHundredNYieldDstPiStatF}{1.7}
\newcommand{\DTwentySixHundredNYieldDstPiSystF}{7.3}
\newcommand{\DTwentySevenFiftyNYield}{23.5}
\newcommand{\DTwentySevenFiftyNYieldStat}{2.1}
\newcommand{\DTwentySevenFiftyNYieldSyst}{5.2}
\newcommand{\DTwentySevenFiftyNMass}{2752.4}
\newcommand{\DTwentySevenFiftyNMassStat}{1.7}
\newcommand{\DTwentySevenFiftyNMassSyst}{2.7}
\newcommand{\DTwentySevenFiftyNWidth}{71}
\newcommand{\DTwentySevenFiftyNWidthStat}{6}
\newcommand{\DTwentySevenFiftyNWidthSyst}{11}
\newcommand{\DTwentySevenFiftyNSig}{4.2}

\newcommand{\DTwentyFourSixtyCYield}{110.8}
\newcommand{\DTwentyFourSixtyCYieldStat}{1.3}
\newcommand{\DTwentyFourSixtyCYieldSyst}{7.5}
\newcommand{\DTwentyFourSixtyCMass}{2465.4}
\newcommand{\DTwentyFourSixtyCMassStat}{0.2}
\newcommand{\DTwentyFourSixtyCMassSyst}{1.1}
\newcommand{\DTwentyFourSixtyCWidth}{50.5}

\newcommand{\DTwentySixHundredCYield}{13.0}
\newcommand{\DTwentySixHundredCYieldStat}{1.3}
\newcommand{\DTwentySixHundredCYieldSyst}{4.5}
\newcommand{\DTwentySixHundredCMass}{2621.3}
\newcommand{\DTwentySixHundredCMassStat}{3.7}
\newcommand{\DTwentySixHundredCMassSyst}{4.2}
\newcommand{\DTwentySixHundredCWidth}{93}
\newcommand{\DTwentySixHundredCSig}{2.8}

\newcommand{\DTwentySevenSixtyCYield}{5.7}
\newcommand{\DTwentySevenSixtyCYieldStat}{0.7}
\newcommand{\DTwentySevenSixtyCYieldSyst}{1.5}
\newcommand{\DTwentySevenSixtyCMass}{2769.7}
\newcommand{\DTwentySevenSixtyCMassStat}{3.8}
\newcommand{\DTwentySevenSixtyCMassSyst}{1.5}
\newcommand{\DTwentySevenSixtyCWidth}{60.9}
\newcommand{\DTwentySevenSixtyCSig}{3.5}

\newcommand{\DTwentyFourSixtyBFRatio}{\ensuremath{1.47 \pm 0.03 \pm 0.16}\xspace}
\newcommand{\DTwentySixHundredBFRatio}{\ensuremath{0.32 \pm 0.02 \pm 0.09}\xspace}
\newcommand{\DTwentySevenSixtyBFRatio}{\ensuremath{0.42 \pm 0.05 \pm 0.11}\xspace}

\newcommand{\BABARPubYear}    {10}
\newcommand{\BABARPubNumber}  {020}

\newcommand{\SLACPubNumber} {14246}
\newcommand{\LANLNumber} {1009.2076}
\newcommand{\lumi}    {454~\invfb}

\begin{document}

\preprint{\babar-PUB-\BABARPubYear/\BABARPubNumber}
\preprint{SLAC-PUB-\SLACPubNumber}

\begin{flushleft}
\ \\
\ \\
\ 
\end{flushleft}
\
\
\begin{flushright}
  \babar-PUB-\BABARPubYear/\BABARPubNumber\\
  SLAC-PUB-\SLACPubNumber \\
  hep-ex/\LANLNumber
\end{flushright}

\title{
{\large \boldmath Observation of new resonances decaying to $D\pi$ and $D^*\pi$ in inclusive $e^+e^-$ collisions near $\sqrt{s}=$10.58 GeV}

}

%% author list as of 01-Jun-2010 (440 authors)
%
\author{P.~del~Amo~Sanchez}
\author{J.~P.~Lees}
\author{V.~Poireau}
\author{E.~Prencipe}
\author{V.~Tisserand}
\affiliation{Laboratoire d'Annecy-le-Vieux de Physique des Particules (LAPP), Universit\'e de Savoie, CNRS/IN2P3,  F-74941 Annecy-Le-Vieux, France}
\author{J.~Garra~Tico}
\author{E.~Grauges}
\affiliation{Universitat de Barcelona, Facultat de Fisica, Departament ECM, E-08028 Barcelona, Spain }
\author{M.~Martinelli$^{ab}$}
\author{A.~Palano$^{ab}$ }
\author{M.~Pappagallo$^{ab}$ }
\affiliation{INFN Sezione di Bari$^{a}$; Dipartimento di Fisica, Universit\`a di Bari$^{b}$, I-70126 Bari, Italy }
% ...
\collaboration{The \babar\ Collaboration}
\noaffiliation

\date{\today}
\begin{abstract}
We present a study of the $D^+\pi^-$, $D^0\pi^+$, and $D^{*+}\pi^-$ systems in inclusive $e^+e^- \rightarrow c\bar{c}$ interactions in a search for new excited $D$ meson states. 
We use a dataset, consisting of $\sim$454 fb$^{-1}$, collected at center-of-mass energies near 10.58 GeV by the 
\mbox{\slshape B\kern-0.1em{\smaller A}\kern-0.1em B\kern-0.1em{\smaller A\kern-0.2em R}} 
detector at the SLAC PEP-II asymmetric-energy collider. 
We observe, for the first time, candidates for the radial excitations of the $D^0$, $D^{*0}$, and $D^{*+}$, as well as the $L=2$ excited states of the $D^0$ and $D^+$, where $L$ is the orbital angular momentum of the quarks.
\end{abstract}


\maketitle

The spectrum of mesons consisting of a charm and an up or a down quark is poorly known. The spectrum of quark-antiquark systems was predicted in 1985 using a relativistic chromodynamic  potential model \cite{Godfrey}.
The low-mass spectrum of the $c~\overline{\kern -0.2em u}$ or $c~\overline{\kern -0.2em d}$ system is comprised of the ground states (1S), the orbital excitations with angular momentum $L$=1,2 (1P,1D), and the first radial excitations (2S). 
In this paper we label the states using the notation $D_J^{(2S+1)}(nL)$, where $J$ is the total angular momentum of the state, $n$ is the radial quantum number, and $L$ and $S$ are the orbital angular momentum and total spin of the quarks.
Besides the ground states ($D,D^*$), only two 1P states, known as the \DTwentyFourTwenty and \DTwentyFourSixty \cite{PDG}, are well-established experimentally since they have relatively narrow widths ($\sim$30 \MeV).
In contrast, the other two 1P states, known as the \DTwentyFourHundred and \DTwentyFourThirty, are very broad ($\sim$300 \MeV), making them difficult to detect \cite{Peskin,BelleBToDstPiPi,BaBarBToDPiPi}.

To search for states not yet observed, we analyze the {\it inclusive} production of the \DPi, \DzPi, and \DstarPi~\cite{complex} final states in the reaction $\epem \rightarrow \cc \rightarrow D^{(*)}\pi X$, where $X$ is any additional system.
We use an event sample consisting of approximately 590 million $\epem\rightarrow\cc$ events (\lumi) produced at \epem center-of-mass (CM) energies near 10.58 GeV and collected with the $\babar$ detector at the SLAC \pep2 asymmetric-energy collider.
Our signal yield for the $L=1$ resonances is more than ten times larger than the best previous study \cite{CDF}, resulting in much greater sensitivity to higher resonances.

The $\babar$ detector is described in detail in Ref.~\cite{babar}.
Charged-particle momenta are measured with a 5-layer, double-sided silicon vertex tracker (SVT) and a 40-layer drift chamber (DCH) inside a 1.5-T superconducting solenoidal magnet.
A calorimeter consisting of 6580 CsI(Tl) crystals is used to measure electromagnetic energy. 
A ring-imaging Cherenkov radiation detector (DIRC), aided by measurements of ionization energy loss, $dE/dx$, in the SVT and DCH, is used for particle identification (PID) of charged hadrons.

The $D \pi$ system is reconstructed in the \DPi and \DzPi modes, where \DCToKPiPi and \DToKPi.
A PID algorithm is applied to all tracks. Charged kaon identification has an average efficiency of 90\% within the acceptance of the detector and an average pion-to-kaon misidentification probability of 1.5\%.

For all channels we perform a vertex fit for the \Dp and \Dz daughters. To improve the signal to background ratio for \DCToKPiPi, we require that the measured flight distance of the \Dp candidate from the \epem interaction region be greater than 5 times its uncertainty.
To improve the signal purity for \DToKPi we require $\cos \theta_{K}> -0.9$ where $\theta_{K}$ is the angle formed by the $K^-$ in the \Dz candidate rest frame with respect to the prior direction of the \Dz~ candidate in the CM reference frame.
The $D \pi$ candidates for both \Dp and \Dz are then reconstructed by performing a vertex fit with an additional charged {\it primary} pion, which originates from the \epem interaction region. For all vertex fits we require a $\chi^2$ probability $>0.1\%$.

In the \DzPi sample, we veto \Dz candidates from $D^{*+}$ or $D^{*0}$ decays by forming $\Dz\pip$ (where the $\pi^+$ is any additional pion in the event) and $\Dz\piz$ combinations, and rejecting the event if the invariant-mass difference between this combination and the $D^0$ candidate is within $2\sigma$ of the nominal $D^*$-$D$ mass difference \cite{PDG}, where $\sigma$ is the detector resolution.

\begin{figure}[t]
%\begin{small}
\includegraphics[clip,width=1.0\columnwidth]{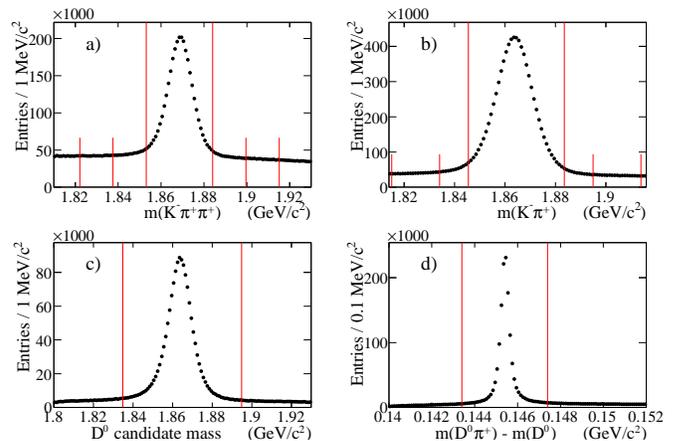}
%\vspace{-0.6cm}
\caption{
 (color online) Mass distribution for a) \Dp and b) \Dz candidates in the \DPi and \DzPi samples.  Plots c) and d) correspond to the \DstarPi sample and show the mass distribution for \Dz candidates and the $\Delta m$ distribution for \Dstar candidates.  The vertical lines show the signal and, in a) and b), the side-band regions.
}
\label{fig:fig1}
%\end{small}
\end{figure}

The \KPiPi and \KPi mass distributions are shown in Figs.~\ref{fig:fig1} a) and \ref{fig:fig1} b). 
We fit these distributions to a linear background and a Gaussian signal; the signal widths obtained are $\sigma_{\Dp}= 6.7$ \mevcc and $\sigma_{\Dz} = 7.6$ \mevcc. 
The signal region is defined to be within $\pm 2.5 \sigma$ of the peak, while sideband regions are defined as the ranges $(\pm 5.0\sigma,\pm 7.5\sigma)$ and $(\pm 4.0\sigma,\pm 6.5\sigma)$ for the \Dp and \Dz, respectively. 
The \Dp signal region has purity $N_S/(N_S+N_B)=65\%$, where $N_S$ ($N_B$) is the number of signal (background) events, while the \Dz~ purity is $83\%$.

The $\Dstar\pim$ system is reconstructed using the \DToKPi and \DToKPiPiPi decay modes. 
A \Dz candidate is accepted if its invariant mass is within 30 \MeVcc of the mean value.
A \Dstar candidate is reconstructed by requiring an additional slow pion ($\pi^+_s$) originating from the \epem interaction region. We select a \Dstar candidate if the mass difference $\Delta m = m(\Km\pip(\pip\pim)\pi^+_s) - m(\Km\pip(\pip \pim))$ is within 2.0 \MeVcc of the mean value.
The \Dz candidate invariant mass distribution and the $\Delta m$ distribution are shown in Figs.~\ref{fig:fig1} c) and \ref{fig:fig1} d). The \Dstar signal purity is $89\%$.
Finally, we reconstruct a $\Dstar\pim$ candidate by combining a \Dstar candidate with an additional charged track identified as a \pim and applying a vertex fit.

Background from $e^+ e^- \to B \bar B$ events, and much of the combinatorial background, are removed by requiring the CM momentum of the $D^{(*)}\pi$ system to be greater than 3.0 \gevc.
In addition, we remove fake primary pion candidates originating mainly from the opposite side of the event by requiring $\cos \theta_{\pi}>-0.8$. The angle $\theta_{\pi}$ is defined in the $D^{(*)}\pi$ rest frame as the angle between the primary pion direction and the prior direction of the $D^{(*)}\pi$ system in the CM frame.

To extract the resonance parameters we define the variables 
$\MDPi = m(\Km \pip \pip \pim) - m(\Km \pip \pip) + m_{\Dp}$ 
and $\MDzPi = m(\Km \pip \pip) - m(\Km \pip) + m_{\Dz}$, 
where m$_{\Dp}$ and m$_{\Dz}$ are the values of the \Dp and \Dz mass \cite{PDG}.
The use of the mass difference improves the resolution on the reconstructed mass to about 3 \MeVcc. 
We remove the contribution due to fake \Dp and \Dz candidates by subtracting the $M(D\pi)$ distributions obtained by selecting events in the \Dp or \Dz candidate mass sidebands.

\begin{figure}[t]
\includegraphics[clip,width=1.\columnwidth]{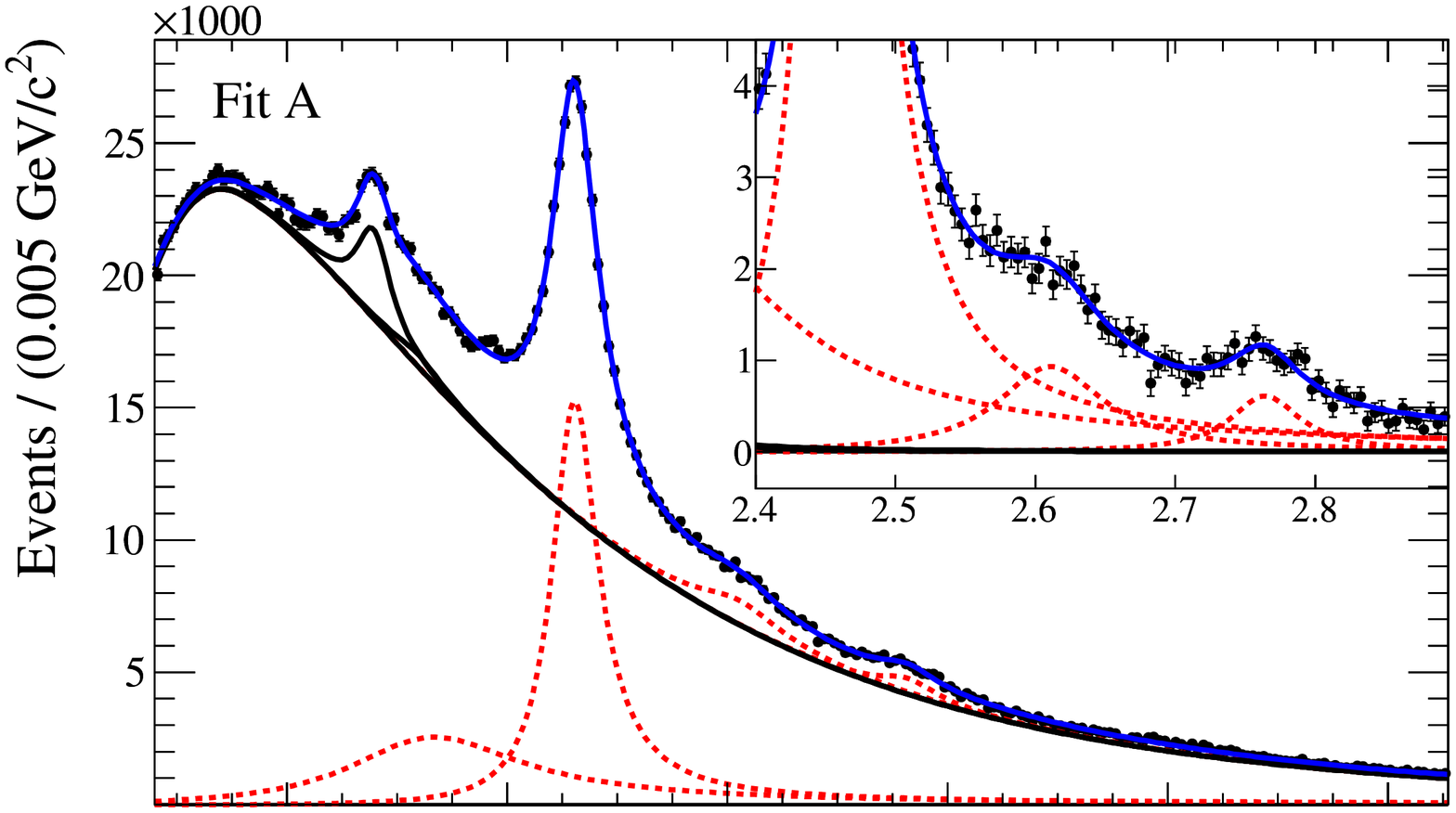}
\includegraphics[clip,width=1.\columnwidth]{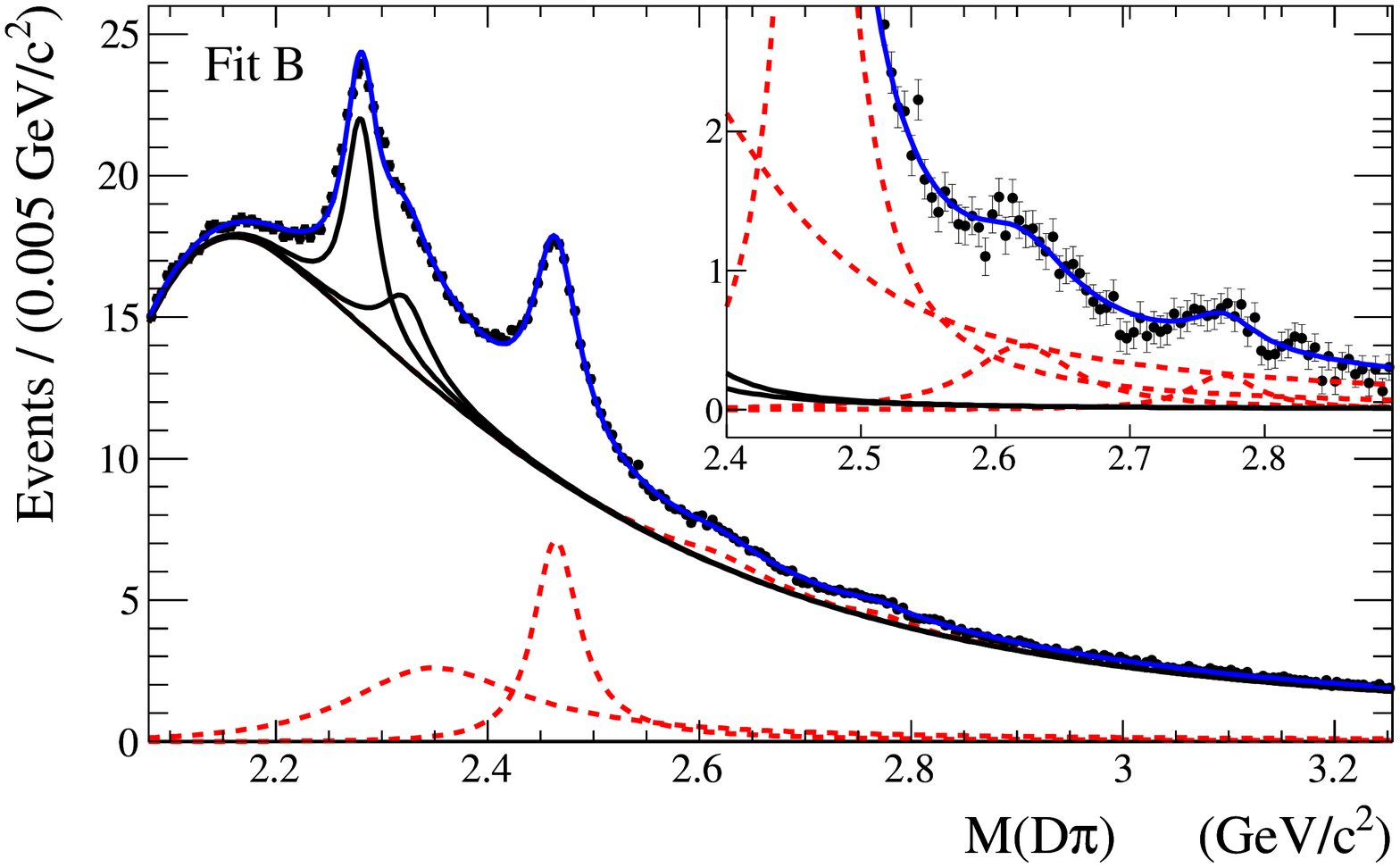}
%\vspace{-0.2cm}
\caption{
(color online) Mass distribution for \DPi (top) and \DzPi (bottom) candidates. Points correspond to data, with the total fit overlaid as a solid curve.
The dotted curves are the signal components.
The lower solid curves correspond to the smooth combinatoric background and to the peaking backgrounds at 2.3 \gevcc.
The inset plots show the distributions after subtraction of the combinatoric background.
}
\label{fig:fig2} 
%\vspace{-0.2cm}
\end{figure}

The \DPi and \DzPi mass spectra are presented in Fig.~\ref{fig:fig2} and show similar features.
\begin{itemize}
\item Prominent peaks for \DTwentyFourSixtyNeutral and \DTwentyFourSixtyCharged .
\item The \DPi mass spectrum shows a peaking background (feeddown) at about 2.3 \gevcc due to decays from the \DTwentyFourTwentyNeutral and \DTwentyFourSixtyNeutral to \DstarPi. The \Dstar in these events decays to $\Dp\pi^0$ and the $\pi^0$ is missing in the reconstruction. The missing $\pi^0$ has very low momentum because the \Dstar decay is very close to threshold. Therefore, these decays have a mass resolution of only 5.8 \MeVcc and a bias of $-143.2$ \MeVcc. Similarly, \DzPi shows peaking backgrounds due to the decays of the \DTwentyFourTwentyCharged and \DTwentyFourSixtyCharged to $\Dstarz\pi^+$ where the \Dstarz decays to $\Dz\pi^0$.
\item Both \DPi and  \DzPi mass distributions show new structures around 2.6 and 2.75 \gevcc. We call these enhancements \DTwentySixHundred and \DTwentySevenSixty.
\end{itemize}

We have compared these mass spectra with those obtained from generic $e^+ e^- \to \bar c c$ Monte Carlo (MC) events.
These events were generated using JETSET \cite{jetset} with all the known particle resonances incorporated.
The events are then reconstructed using a detailed GEANT4 \cite{geant4} detector simulation and the event selection procedure used for the data.
In addition, we study $D\pi$ mass spectra from the \Dp and \Dz candidate mass sidebands, as well as mass spectra for wrong-sign $\Dp\pip$ and $\Dz\pim$ samples.
We find no backgrounds or reflections that can cause the structures at 2.6 and 2.76 \gevcc.
In the study of the \DzPi final state we find a peaking background due to events where the \Dz candidate is not a true \Dz, but the \Km candidate and the primary \pip candidate are from a true \DToKPi decay.
These combinations produce enhancements in \MDzPi both in the \Dz candidate mass signal region and sidebands. However, this background is linear as a function of the \Dz candidate mass, is removed by the sideband subtraction.

The smooth background is modeled using the function:
\begin{equation}
B(x) =  P(x) \times \begin{cases} 
               e^{c_1x+c_2x^2}  & \text{ for $x \leq x_0$, }\\
               e^{d_0+d_1x+d_2x^2} & \text{ for $x>x_0$, }
              \end{cases}
\label{eq:background}
\end{equation}
where $P(x) \equiv {\frac{1}{2x}}\sqrt{[x^2-(m_{D}+m_{\pi})^2][x^2-(m_{D}-m_{\pi})^2]}$ is a two-body phase-space factor and $x = M(D\pi)$. Only four parameters are free in the piece-wise exponential: $c_1,\ c_2,\ d_2,$ and $x_0$. 
The parameters $d_0$ and $d_1$ are fixed by requiring that $B(x)$ be continuous and differentiable at the transition point $x_0$. 
We account for the feeddown of peaking backgrounds by convolving Breit-Wigner (BW) functions \cite{BreitWigner} with a function describing the resolution and bias obtained from the simulation of these decays. The mass and width of the \DTwentyFourTwenty feeddown are fixed to the values obtained in the \DstarPi analysis described below, while the parameters of the \DTwentyFourSixty feeddown are fixed  to those of the true \DTwentyFourSixty in the same $M(D\pi)$ distribution.

The \DTwentyFourSixty is modeled using a relativistic BW function with the appropriate Blatt-Weisskopf centrifugal barrier factor~\cite{PDG}. The \DTwentySixHundred and \DTwentySevenSixty are modeled with relativistic BW functions~\cite{PDG}.
Finally, although not visible in the \MDPi mass distribution, we include a BW function to account for the known resonance \DTwentyFourHundred, which is expected to decay to this final state. 
The $\chi^2$ per number of degrees of freedom (NDF) of the fit decreases from 596/245 to 281/242 when this resonance is included.
This resonance is very broad and is present together with the feeddown and \DTwentyFourSixtyNeutral; therefore we restrict its mass and width parameters to be within 2$\sigma$ of the known values \cite{BaBarBToDPiPi}. 
The shapes of the signal components are corrected for a small variation of the efficiency as a function of $M(D\pi)$ and are multiplied by the two-body phase-space factor. 
They are also corrected for the mass resolution by convolving them with the resolution function determined from MC simulation of signal decays. 
The fit to the \MDPi distribution (Fit A) is shown in Fig.~\ref{fig:fig2} (top).
The results of this fit, as well as fits to the other final states described below, are shown in Table~\ref{tab:summarypars}.
In this table, the significance for each new signal is defined as the signal yield divided by its total uncertainty.

\begin{figure}[t]

\includegraphics[clip,width=1.\columnwidth]{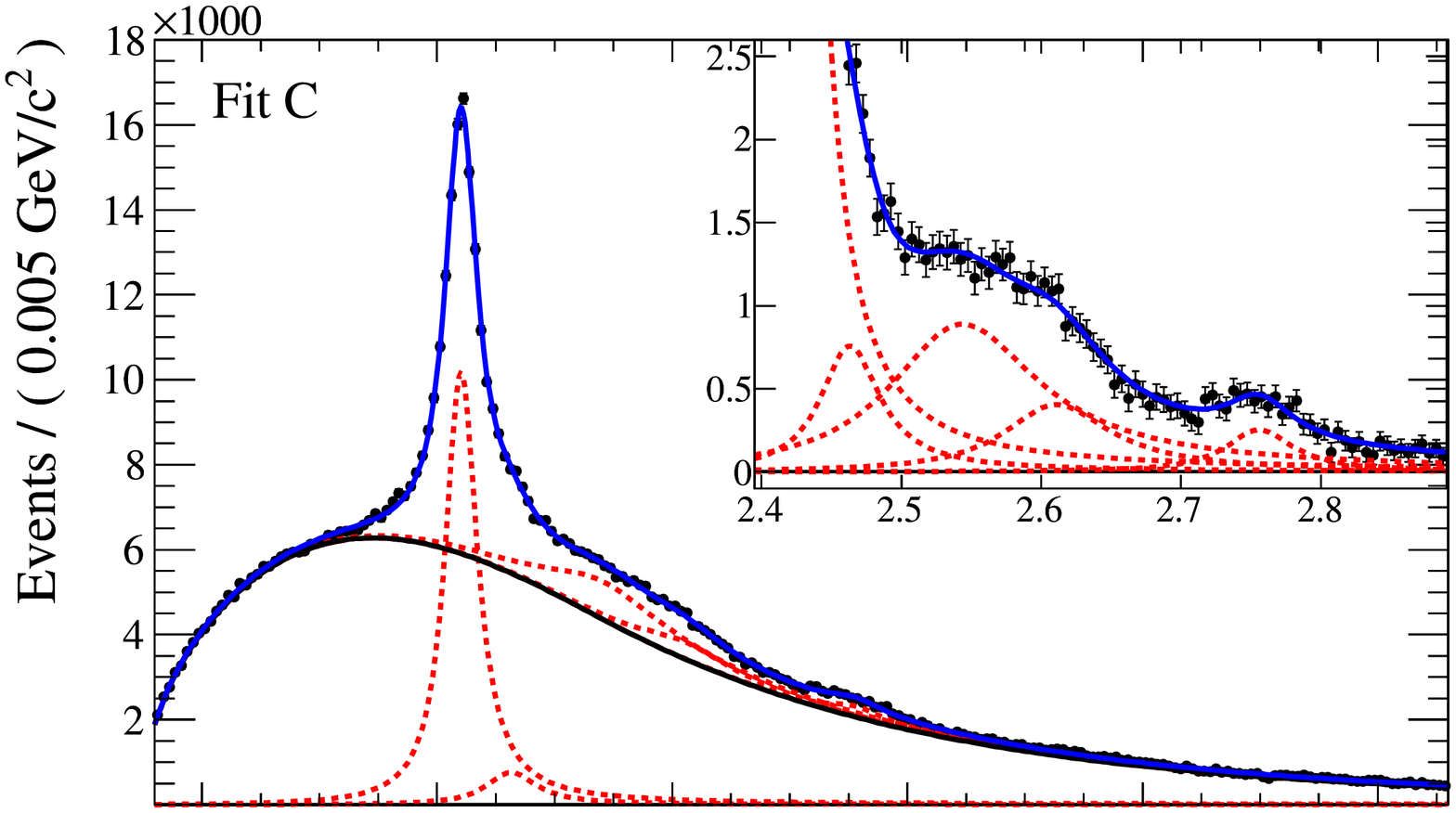}
\includegraphics[clip,width=1.\columnwidth]{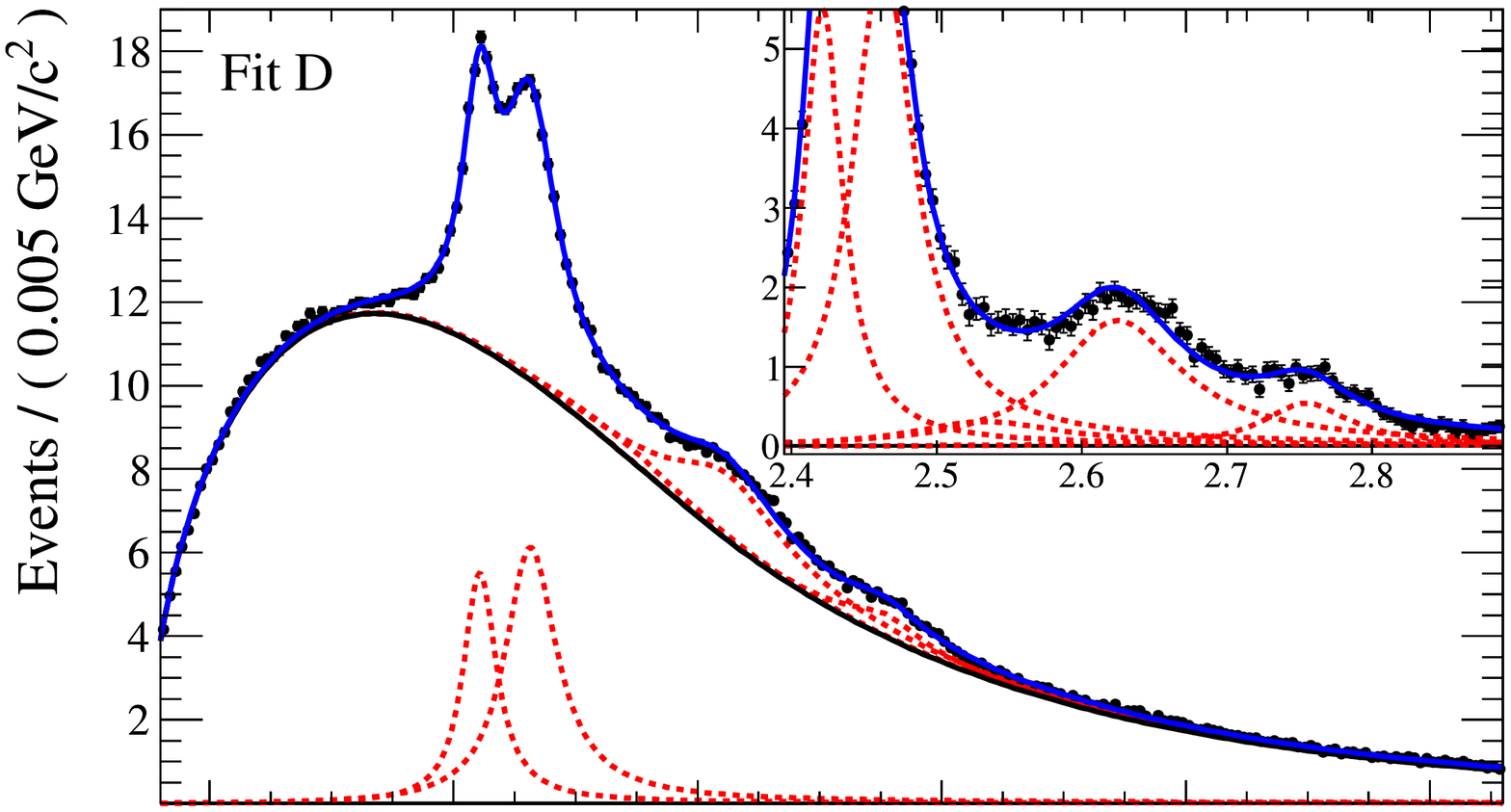}
\includegraphics[clip,width=1.\columnwidth]{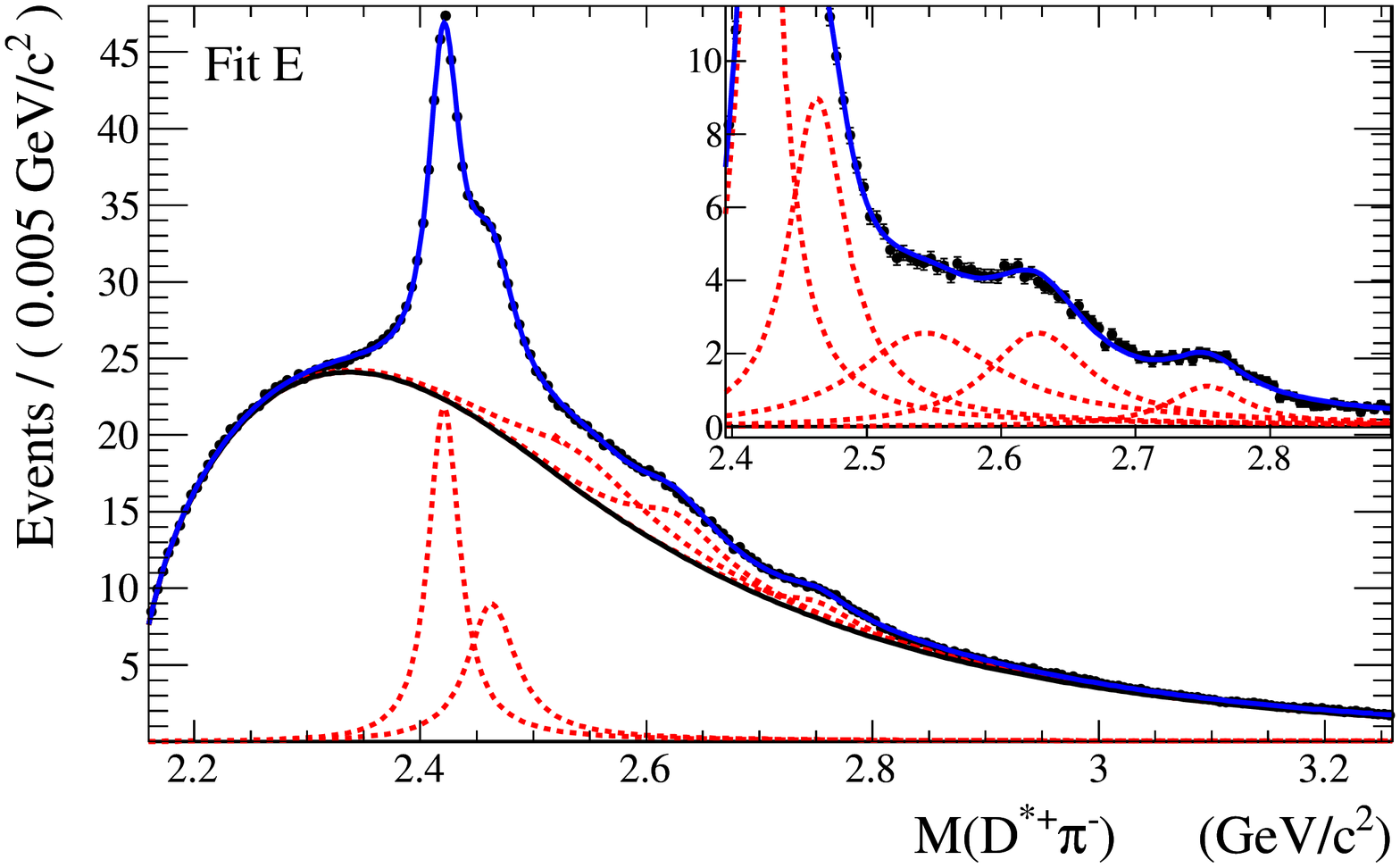}

%\vspace{-0.3cm}
\caption{
(color online) Mass distributions for \DstarPi candidates. Top: candidates with $|\cosHelicity|>0.75$. Middle: candidates with $|\cosHelicity|<0.5$. Bottom: all candidates. Points correspond to data, with the total fit overlaid as a solid curve. 
The lower solid curve is the combinatoric background, and the dotted curves are the signal components.
The inset plots show the distributions after subtraction of the combinatoric background.
}
\label{fig:fig2b} 
%\vspace{-0.2cm}
\end{figure}

The fit to the \DzPi mass spectrum is similar to that described for the \DPi system.
Because the feeddown is larger and the statistical precision of the resonances is not as good as for \DPi, we fix the width parameters of all resonances to the values determined from \DPi assuming isospin symmetry. 
The fit to the \MDzPi mass distribution (Fit B) is shown in Fig.~\ref{fig:fig2} (bottom); this fit has $\chi^2/$NDF of 278/224.
We find consistent mass values for both \DTwentySixHundred and \DTwentySevenSixty in the fits of the \DPi and \DzPi mass distributions.

\begin{table*}[t]

\caption[Summary parameters]{Summary of the results. The first error is statistical and the second is systematic; ``fixed'' indicates the parameters were fixed to the values from Fit A or C. The significance is defined as the yield divided by its total error.
}  
\label{tab:summarypars}

\renewcommand{\arraystretch}{1.10}

\begin{tabular*}{\linewidth}{
@{\extracolsep{\fill}}l
@{\extracolsep{\fill}}l
@{\extracolsep{\fill}}c
@{\extracolsep{\fill}}c
@{\extracolsep{\fill}}c
@{\extracolsep{\fill}}c
@{\extracolsep{\fill}}c
}\hline\hline

{\bf Resonance} & {\bf Channel(Fit) } & {\bf Efficiency (\%)} & {\bf Yield (x$10^3$)}  & {\bf Mass (\mevcc)}  &  {\bf Width (\mev)} & {\bf Significance} \\ 
\hline   
\hline

\DTwentyFourTwentyNeutral & \DstarPi(C) &  & \DTwentyFourTwentyNYield $\pm$\DTwentyFourTwentyNYieldStat $\pm$\DTwentyFourTwentyNYieldSyst & \DTwentyFourTwentyNMass $\pm$\DTwentyFourTwentyNMassStat $\pm$\DTwentyFourTwentyNMassSyst &  \DTwentyFourTwentyNWidth $\pm$\DTwentyFourTwentyNWidthStat $\pm$\DTwentyFourTwentyNWidthSyst& \\
                          & \DstarPi(E) & $1.09\pm0.03$ & \DTwentyFourTwentyNYieldF $\pm$\DTwentyFourTwentyNYieldStatF $\pm$\DTwentyFourTwentyNYieldSystF & \DTwentyFourTwentyNMass(fixed) & \DTwentyFourTwentyNWidth(fixed) & \\
\hline

\DTwentyFourSixtyNeutral & \DPi(A) & $1.29\pm0.03$ & \DTwentyFourSixtyNYield $\pm$\DTwentyFourSixtyNYieldStat $\pm$\DTwentyFourSixtyNYieldSyst & \DTwentyFourSixtyNMass $\pm$\DTwentyFourSixtyNMassStat $\pm$\DTwentyFourSixtyNMassSyst & \DTwentyFourSixtyNWidth $\pm$\DTwentyFourSixtyNWidthStat $\pm$\DTwentyFourSixtyNWidthSyst & \\
                  & \DstarPi(E) & $ 1.12\pm0.04$ & \DTwentyFourSixtyNYieldDstPi  $\pm$\DTwentyFourSixtyNYieldDstPiStat $\pm$\DTwentyFourSixtyNYieldDstPiSyst &  \DTwentyFourSixtyNMass(fixed) & \DTwentyFourSixtyNWidth(fixed)& \\
\hline

\DTwentyFiveFiftyNeutral & \DstarPi(C) &   & \DTwentyFiveFiftyNYield $\pm$\DTwentyFiveFiftyNYieldStat $\pm$\DTwentyFiveFiftyNYieldSyst & \DTwentyFiveFiftyNMass $\pm$\DTwentyFiveFiftyNMassStat $\pm$\DTwentyFiveFiftyNMassSyst & \DTwentyFiveFiftyNWidth $\pm$\DTwentyFiveFiftyNWidthStat $\pm$\DTwentyFiveFiftyNWidthSyst & \DTwentyFiveFiftyNSig$\sigma$ \\
                         & \DstarPi(E) & $1.14\pm0.04$  & \DTwentyFiveFiftyNYieldF $\pm$\DTwentyFiveFiftyNYieldStatF $\pm$\DTwentyFiveFiftyNYieldSystF & \DTwentyFiveFiftyNMass(fixed) & \DTwentyFiveFiftyNWidth(fixed) & \\
\hline

\DTwentySixHundredNeutral  & \DPi(A) & $1.35\pm0.05$ & \DTwentySixHundredNYield $\pm$\DTwentySixHundredNYieldStat $\pm$\DTwentySixHundredNYieldSyst  & \DTwentySixHundredNMass $\pm$\DTwentySixHundredNMassStat  $\pm$\DTwentySixHundredNMassSyst & \DTwentySixHundredNWidth $\pm$\DTwentySixHundredNWidthStat$\pm$\DTwentySixHundredNWidthSyst & \DTwentySixHundredNSig$\sigma$\\
                           & \DstarPi(D) &  & \DTwentySixHundredNYieldDstPi $\pm$\DTwentySixHundredNYieldDstPiStat $\pm$\DTwentySixHundredNYieldDstPiSyst & \DTwentySixHundredNMass(fixed) & \DTwentySixHundredNWidth(fixed) & \DTwentySixHundredNDstPiSig$\sigma$\\
                           & \DstarPi(E) & $1.18\pm0.05$ & \DTwentySixHundredNYieldDstPiF $\pm$\DTwentySixHundredNYieldDstPiStatF $\pm$\DTwentySixHundredNYieldDstPiSystF & \DTwentySixHundredNMass(fixed) & \DTwentySixHundredNWidth(fixed)& \\
\hline

\DTwentySevenFiftyNeutral  & \DstarPi(E) & $ 1.23\pm0.07$  & \DTwentySevenFiftyNYield $\pm$\DTwentySevenFiftyNYieldStat $\pm$\DTwentySevenFiftyNYieldSyst & \DTwentySevenFiftyNMass $\pm$\DTwentySevenFiftyNMassStat $\pm$\DTwentySevenFiftyNMassSyst & \DTwentySevenFiftyNWidth $\pm$\DTwentySevenFiftyNWidthStat $\pm$\DTwentySevenFiftyNWidthSyst & \DTwentySevenFiftyNSig$\sigma$\\
\hline

\DTwentySevenSixtyNeutral  & \DPi(A) &  $ 1.41\pm0.09$ & \DTwentySevenSixtyNYield $\pm$\DTwentySevenSixtyNYieldStat $\pm$\DTwentySevenSixtyNYieldSyst  & \DTwentySevenSixtyNMass $\pm$\DTwentySevenSixtyNMassStat $\pm$\DTwentySevenSixtyNMassStat & \DTwentySevenSixtyNWidth $\pm$\DTwentySevenSixtyNWidthStat $\pm$\DTwentySevenSixtyNWidthSyst & \DTwentySevenSixtyNSig$\sigma$\\
\hline

\DTwentyFourSixtyCharged   & \DzPi(B) &   &  \DTwentyFourSixtyCYield $\pm$\DTwentyFourSixtyCYieldStat $\pm$\DTwentyFourSixtyCYieldSyst & \DTwentyFourSixtyCMass $\pm$\DTwentyFourSixtyCMassStat $\pm$\DTwentyFourSixtyCMassSyst & \DTwentyFourSixtyCWidth(fixed) & \\
\hline

\DTwentySixHundredCharged  & \DzPi(B) &   & \DTwentySixHundredCYield $\pm$\DTwentySixHundredCYieldStat $\pm$\DTwentySixHundredCYieldSyst  & \DTwentySixHundredCMass $\pm$\DTwentySixHundredCMassStat $\pm$\DTwentySixHundredCMassSyst & \DTwentySixHundredCWidth(fixed) & \DTwentySixHundredCSig$\sigma$\\
\hline 

\DTwentySevenSixtyCharged  & \DzPi(B) &   & \DTwentySevenSixtyCYield $\pm$\DTwentySevenSixtyCYieldStat $\pm$\DTwentySevenSixtyCYieldSyst &  \DTwentySevenSixtyCMass $\pm$\DTwentySevenSixtyCMassStat $\pm$\DTwentySevenSixtyCMassSyst & \DTwentySevenSixtyCWidth(fixed) & \DTwentySevenSixtyCSig$\sigma$\\
\hline 

\hline        
\end{tabular*}
\end{table*}

We now search for these new states in the \DstarPi decay mode. 
We define the variable 
$\MDstarPi = m(\Km \pip (\pip \pim) \pi^+_s \pim) - m(\Km \pip (\pip \pim) \pi^+_s) + m_{\Dstar}$ 
where $m_{\Dstar}$ is the value of the $\Dst$ mass~\cite{PDG}. 
The \DstarPi mass distribution is shown in Fig.~\ref{fig:fig2b} and shows the following features.
\begin{itemize}
\item Prominent \DTwentyFourTwentyNeutral and \DTwentyFourSixtyNeutral peaks.
\item Two additional enhancements at $\sim$2.60 \GeVcc and $\sim$2.75 \GeVcc, which we initially denote as \DTwentySixHundredNeutral and \DTwentySevenFiftyNeutral.
\end{itemize} 
Studies of the generic MC simulation as well as studies of the \Dstar sidebands and the wrong-sign sample (\DstarPiWS) show no peaking backgrounds in this mass spectrum.

We fit \MDstarPi by parametrizing the background with the function in Eq.~(\ref{eq:background}). 
The \DTwentyFourTwentyNeutral and \DTwentyFourSixtyNeutral resonances are modeled using relativistic BW functions with appropriate Blatt-Weisskopf form factors. 
The \DTwentySixHundredNeutral and \DTwentySevenFiftyNeutral are modeled with relativistic BW functions.
The broad resonance \DTwentyFourThirtyNeutral is known to decay to this final state, however, this fit is insensitive to it due to its large width ($\sim$380 \MeV) \cite{BelleBToDstPiPi} and because the background parameters are free.

\begin{table}[b] 
%\vspace{-0.7cm}
\caption[theory values]{Properties of the predicted states \cite{Godfrey}.
The value of the parameter $h$ depends on the state.
}  
\label{tab:theory}
%\vspace{-.3cm}
\begin{center}

\begin{tabular*}{\columnwidth}{
@{\extracolsep{\fill}}l
@{\extracolsep{\fill}}c
@{\extracolsep{\fill}}c
@{\extracolsep{\fill}}c
}
\hline
      State & Predicted Mass  & $J^P$  & \cosHelicity Distribution \\  
\hline
$D^1_{0}(2S)$   & 2.58 \GeVcc & $0^{-}$  & $\propto \cos^2\theta_H$ \\
$D^{3}_{1}(2S)$ & 2.64 \GeVcc & $1^{-}$  & $\propto \sin^2\theta_H$ \\
$D^1_{1}(1P)$   & 2.44 \GeVcc & $1^{+}$  & $\propto 1+h\cos^2\theta_H$ \\
$D^{3}_{0}(1P)$ & 2.40 \GeVcc & $0^{+}$  &  decay not allowed \\
$D^{3}_{1}(1P)$ & 2.49 \GeVcc & $1^{+}$  & $\propto 1+h\cos^2\theta_H$ \\
$D^{3}_{2}(1P)$ & 2.50 \GeVcc & $2^{+}$  & $\propto \sin^2\theta_H$ \\
$D^1_{2}(1D)$   & $\sim$2.83 \GeVcc & $2^{-}$  & $\propto 1+h\cos^2\theta_H$ \\
$D^{3}_{1}(1D)$ & 2.82 \GeVcc & $1^{-}$  & $\propto \sin^2\theta_H$ \\
$D^{3}_{2}(1D)$ & $\sim$2.83 \GeVcc & $2^{-}$  & $\propto 1+h\cos^2\theta_H$ \\
$D^{3}_{3}(1D)$ & 2.83 \GeVcc & $3^{-}$  & $\propto \sin^2\theta_H$ \\
   \hline 
   \end{tabular*} 
\end{center}
\end{table}

Due to the vector nature of the \Dstar, the \DstarPi final state contains additional information about the spin-parity ($J^P$) quantum numbers of the resonances. 
In the rest frame of the \Dstar, we define the {\it helicity} angle $\theta_H$ as the angle between the primary pion $\pi^-$ and the slow pion $\pi^+$ from the \Dstar decay. 
The distributions in \cosHelicity~ for the predicted resonances, assuming parity conservation, are given in Table~\ref{tab:theory}.
Initially, we have attempted to fit the \MDstarPi distribution incorporating only two new signals at $\sim$2.6 \GeVcc and at $\sim$2.75 \GeVcc. 
However, when we extract the yields as a function of \cosHelicity~ we find that the mean value of the peak at $\sim$2.6 \GeVcc increases by $\sim$70 \MeVcc between $\cosHelicity=-1$ and $\cosHelicity=0$, and decreases again as $\cosHelicity\rightarrow{+1}$. 
This behaviour suggests two resonances with different helicity-angle distributions are present in this mass region.
To proceed we incorporate a new component, which we call \DTwentyFiveFiftyNeutral, into our model at $\sim$2.55 \GeVcc. 
We extract the parameters of this component by requiring $|\cosHelicity|>0.75$ in order to suppress the other resonances.
In this fit (Fit C), shown in Fig.~\ref{fig:fig2b} (top), we fix the parameters of the \DTwentyFourSixtyNeutral and \DTwentySixHundredNeutral to those measured in \DPi. We obtain a $\chi^2/$NDF of 214/205 for this fit.
This fit also determines the parameters of the \DTwentyFourTwentyNeutral. 
We then perform a complementary fit (Fit D), shown in Fig.~\ref{fig:fig2b} (middle), in which we require $|\cosHelicity|<0.5$ to discriminate in favor of the \DTwentySixHundredNeutral. We obtain a $\chi^2/$NDF of 210/209 for this fit.
To determine the final parameters of the \DTwentySevenFiftyNeutral signal we fit the total \DstarPi sample while fixing the parameters of all other BW components to the values determined in the previous fits. 
This final fit (Fit E), shown in Fig.~\ref{fig:fig2b} (bottom), has a $\chi^2/$NDF of 244/207. 

\begin{figure*}[ht!]
\begin{center}
\includegraphics[clip,width=2.05\columnwidth]{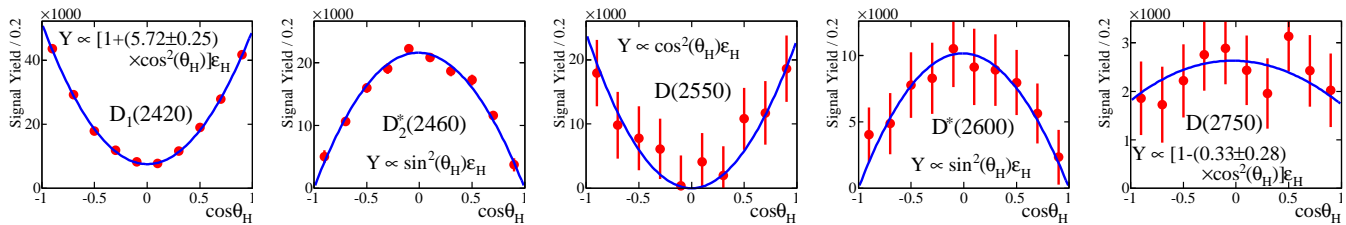}
\end{center}
%\vspace{-0.7cm}
\caption{ 
(color online) Distribution in \cosHelicity~ for each signal in \DstarPi. The error bars include statistical and correlated systematic uncertainties. The curve is a fit using the function Y shown in the plot; $\varepsilon_H$ is the efficiency as a function of \cosHelicity.
}
\label{fig:fig3}
%\vspace{-0.5cm}
\end{figure*}

Systematic uncertainties on all fit results are estimated by varying the parameters that were fixed in the fits and by varying the bin width and mass range of the histograms. 
In addition, the BW shape used for the new signals is replaced by that for a D-wave decay, and we vary the background model according to deviations observed when this model is used to fit the smooth distribution in the wrong-sign samples. 
A systematic uncertainty is also estimated from a possible contribution of the \DTwentyFourThirty. 
Finally, we estimate uncertainties on the mass values due to uncertainties in the magnetic field and the SVT material density.
Effects due to possible interference between the decay amplitudes for different excited states and the background amplitudes are ignored in this inclusive analysis.

The final model for the \MDstarPi distribution is used to extract the signal yields as a function of \cosHelicity. We divide the data into 10 sub-samples corresponding to \cosHelicity intervals of 0.2 between $-1$ and $+1$. Each sample is fitted with all shape parameters fixed to the values determined above. The yields extracted from these fits are plotted for each signal in Fig.~\ref{fig:fig3}.
For the \DTwentyFourTwenty we measure the helicity parameter $h=5.72\pm0.25$, where the error includes both statistical and systematic uncertainties. This value is consistent with the measurement by ZEUS \cite{ZEUS}. The \cosHelicity distributions of the \DTwentyFourSixty and \DTwentySixHundred are consistent with the expectations for {\it natural parity}, defined by $P=(-1)^J$, and leading to a $\sin^2{\theta_H}$ distribution. This observation supports the assumption that the enhancement assigned to the \DTwentySixHundred in the \DPi and \DstarPi belong to the same state; only states with natural parity can decay to both \DPi and \DstarPi.
The \cosHelicity~ distribution for the \DTwentyFiveFiftyNeutral is consistent with pure $\cos^2{\theta_H}$ as expected for a $J^P=0^-$ state.

The ratio of branching fractions $\frac{B(\Dstst\rightarrow\DPi)}{B(\Dstst\rightarrow\DstarPi)}$ (where \Dstst labels any resonance) can be useful in the identification of the new signals with predicted states. 
We compute this ratio for the \DTwentyFourSixtyNeutral, \DTwentySixHundredNeutral, and \DTwentySevenFiftyNeutral using the yields obtained from the fits to the total samples and correcting for the reconstruction efficiency:
$( N_{D\pi}/\varepsilon_{D\pi}) / (N_{D^*\pi} / \varepsilon_{D^*\pi} )$.
The efficiencies and yields are shown in Table~\ref{tab:summarypars}. We find the following ratios: 
\begin{align*}
&\frac{B(\DTwentyFourSixtyNeutral\rightarrow\DPi)}{B(\DTwentyFourSixtyNeutral\rightarrow\DstarPi)}=\DTwentyFourSixtyBFRatio ,\\
&\frac{B(\DTwentySixHundredNeutral\rightarrow\DPi)}{B(\DTwentySixHundredNeutral\rightarrow\DstarPi)}=\DTwentySixHundredBFRatio,\\
&\frac{B(\DTwentySevenSixtyNeutral\rightarrow\DPi)}{B(\DTwentySevenFiftyNeutral\rightarrow\DstarPi)}=\DTwentySevenSixtyBFRatio.
\end{align*}
The first uncertainty is due to the statistical uncertainty on the yields. 
The second uncertainty includes the systematic uncertainty on the yields, the systematic uncertainty due to differences in PID and tracking efficiency, and the errors from the branching fractions for the decay chains~\cite{PDG}. 
Although in the last ratio the signal in the numerator may not be the same as the signal in the denominator, we determine the ratio, as it may help elucidate the nature of this structure.

In summary, we have analyzed the inclusive production of the \DPi, \DzPi, and \DstarPi systems in search of new $D$ meson resonances using \lumi of data collected by the \babar~ experiment. 
We observe for the first time four signals, which we denote \DTwentyFiveFiftyNeutral, \DTwentySixHundredNeutral, \DTwentySevenFiftyNeutral,  and \DTwentySevenSixtyNeutral. 
We also observe the isospin partners \DTwentySixHundredCharged and \DTwentySevenSixtyCharged. 
The \DTwentyFiveFiftyNeutral and \DTwentySixHundredNeutral have mass values and \cosHelicity distributions that are consistent with the predicted radial excitations $D^1_0(2S)$ and $D_1^{3}(2S)$. 
The \DTwentySevenSixtyNeutral signal observed in \DPi is very close in mass to the \DTwentySevenFiftyNeutral signal observed in \DstarPi; however, their mass and width values differ by 2.6$\sigma$ and 1.5$\sigma$, respectively.
Four $L=2$ states are predicted to lie in this region \cite{Godfrey}, but only two are expected to decay to \DPi.
This may explain the observed features.

We are grateful for the excellent luminosity and machine conditions
provided by our \pep2\ colleagues, 
and for the substantial dedicated effort from
the computing organizations that support \babar.
The collaborating institutions wish to thank 
SLAC for its support and kind hospitality. 
This work is supported by
DOE
and NSF (USA),
NSERC (Canada),
CEA and
CNRS-IN2P3
(France),
BMBF and DFG
(Germany),
INFN (Italy),
FOM (The Netherlands),
NFR (Norway),
MES (Russia),
MICIIN (Spain),
STFC (United Kingdom). 
Individuals have received support from the
Marie Curie EIF (European Union),
the A.~P.~Sloan Foundation (USA)
and the Binational Science Foundation (USA-Israel).

\bibliographystyle{unsrt}

\begin{thebibliography}{99}

\bibitem{Godfrey} S. Godfrey and N. Isgur, Phys.~Rev.~D {\bf 32}, 189 (1985).

\bibitem{PDG} C. Amsler \etal (Particle Data Group), Phys.~Lett.~B {\bf 667}, 1 (2008).
 
\bibitem{Peskin} A. F. Falk and M. E. Peskin, Phys.~Rev.~D {\bf 49}, 3320 (1994).

\bibitem{BelleBToDstPiPi} K. Abe \etal(BELLE collaboration),  Phys.~Rev.~D {\bf 69}, 112002 (2004).

\bibitem{BaBarBToDPiPi} B. Aubert \etal(\babar~ collaboration),  Phys.~Rev.~D {\bf 79}, 112004 (2009).

\bibitem{complex}Charge conjugates are implied throughout this paper. 

\bibitem{CDF}A. Abulencia \etal(CDF collaboration), Phys.~Rev.~D {\bf 73}, 051104 (2006).

\bibitem{babar}B. Aubert \etal(\babar~ collaboration), Nucl. Instrum. Methods in Phys. Res. Sect. A  {\bf 479}, 1 (2002).

\bibitem{jetset}  T. Sj\"{o}strand, Computer Physics Commun. {\bf 82}, 74 (1994).

\bibitem{geant4} S. Agostinelli \etal(GEANT4 collaboration), Nucl. Instrum. Methods in Phys. Res. Sect. A {\bf 506}, 250 (2003).

\bibitem{BreitWigner} G. Breit and E. Wigner, Phys.~Rev. {\bf 49}, 519 (1936).

\bibitem{ZEUS}S. Chekanov \etal(ZEUS collaboration), Eur. Phys. J. C {\bf 60}, 25 (2009).

\end{thebibliography}

\end{document}